\documentclass[twocolumn]{article}
\usepackage{nolta2014}
\usepackage{amsmath}
\usepackage{txfonts}
\usepackage{graphicx}
\usepackage{tikz}
\usetikzlibrary{trees}
\usepackage{color}
\usepackage{paralist}
\usepackage{dsfont}
\usetikzlibrary{shapes,snakes}
\usepackage{subfig}
\usepackage{wrapfig}
\usepackage{float}
\usepackage{blindtext}

\begin{document}

\title{Image-based Parameter Inference for \\ Spatio-temporal models of Organogenesis}

\author{Britta Velten, Erkan Uenal and Dagmar Iber}

\address{Department for Biosystems Science and Engineering, ETH Zurich, Mattenstrasse 26, 4058 Basel, Switzerland; \\
Swiss Institute of Bioinformatics (SIB), Switzerland\\
Email: britta.velten@arcor.de, dagmar.iber@bsse.ethz.ch
}

\maketitle

\abstract
Advances in imaging technology now provide us with detailed 3D data on gene expression patterns in developing embryos. This information can be used to build predictive mathematical models of embryogenesis. Current modelling approaches are, however, limited by lack of methods to automatically infer the regulatory networks and the parameter values from the image-based information. Here we make a first step to the development of such methods. We use limb bud development as a model system. For a given regulatory network we developed a decision tree based algorithm to automatically determine parameter values for which the model reproduces the expression patterns. Starting from this parameter set, local optimization was performed to further reduce the chosen goodness-of-fit measure. This approach allowed us to recover the target expression patterns, as judged by eye, and thus provides a first step towards the automated inference of parameter values for a given regulatory network. 
\endabstract

\section{Introduction}

Developmental processes are controlled by complex regulatory networks. Decades of genetic experiments have defined the core regulatory proteins for most developmental processes and many regulatory links. The resulting regulatory networks are complex and the regulatory interactions change dynamically as the embryo is developing. As a consequence, our understanding of the regulatory logic that controls patterning in time and space remains limited. Mathematical modelling offers the opportunity to integrate the experimental information into a consistent framework and to define the underlying regulatory mechanisms \cite{Iber:CurrOpinGenetDev:2012}.

\begin{figure}[t!]
\centering
\includegraphics[width=\columnwidth]{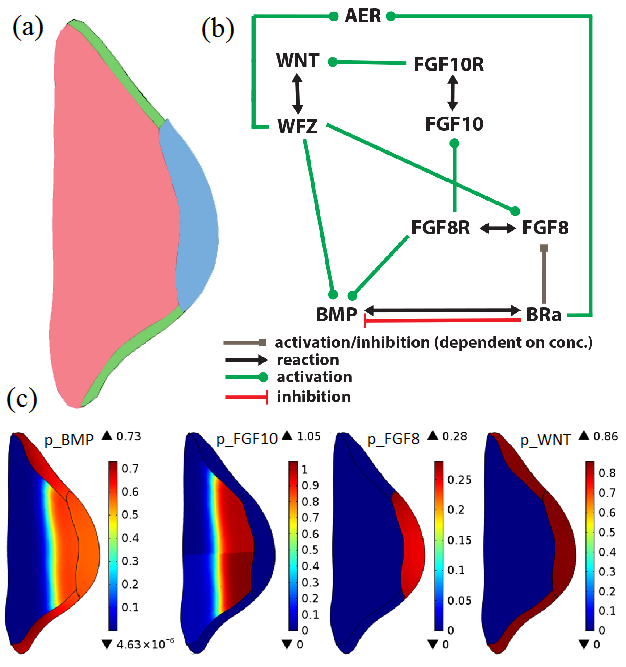}
\caption{The Limb Bud Model System. (a) Subdomains of the limb bud: mesenchyme (red), apical ectodermal ridge AER (blue) and additional ectoderm (green). (b)  The core network of regulatory interactions. (c) \textit{In silico} generated expression patterns using parameter values in table 1, representing the experimentally observed patterns.}
\label{intro}
\end{figure}

To obtain a mathematical model with predictive value the model must be firmly rooted in experimental data. Available experimental data mainly consists of images that show the spatial distribution of mRNAs as a measure of gene expression at the different stages of development. This information is not quantitative, but provides a qualitative indication of expression patterns. Current models of the spatio-temporal processes in the embryo are largely hand-tuned  to reproduce the experimental data both in wild type and mutants \cite{Probst:Development:2011}. Computational methods are largely missing for the image-based inference of the the biological network architecture and the parameter values, even though methods for the estimation of parameter values for partial differential equation (PDE) models have been established. An important limitation are the higher computational costs for the simulations, which renders many approaches computationally infeasible for models of organogenesis. Here, we will focus on a model for mouse limb bud development to illustrate the challenges in inferring parameter values from the available data.

Limb buds grow out of the flank at about day 9 of mouse embryonic development \cite{Zeller:NatRevGenet:2009}. Importantly, the limb bud consists of different tissue domains, and the expression of some of the proteins is restricted to particular subdomains, i.e. to either the mesenchyme (Fig.\ref{intro}a, red part), the apical ectodermal ridge (AER) (Fig.\ref{intro}a, blue part) or the ectoderm (Fig.\ref{intro}a, green and blue part). We will focus on the early regulatory interactions, and we will thus restrict ourselves to the core regulatory interactions between fibroblastic growth factor (FGF) 10, WNT3, FGF8 and BMP. Decades of experiments have defined the core regulatory interactions (Fig.\ref{intro}b). Thus, WNT2B from the flank induces the expression of \textit{Fgf10} in the mesenchyme. FGF10 signalling induces the expression of \textit{Wnt3} in the ectoderm. WNT signalling is necessary for the development of the AER, which expresses \textit{Fgf8} and other \textit{Fgfs}. FGF8 diffuses into the mesenchyme and maintains the expression of \textit{Fgf10}. FGF8 together with WNT3 also induce the expression of \textit{Bmp}s and BMPs supports the development of the AER.

In the following, we will build a mathematical model that represents these regulatory interactions. The focus will then be on the inference of the parameter values. To test the approach we will use simulated \textit{in silico} data rather than experimental gene expression data as obtained from \textit{in situ} hybridisation.

\section{Results \& Discussion}

\subsection{The Model}
To keep the computational costs to a minimum we will limit ourselves to a 2D limb bud domain (Fig \ref{intro}a). To simulate the regulatory network (Fig.\ref{intro}b) with $n$ components on a limb bud domain we use a set of coupled partial differential equations of reaction-diffusion type, i.e. 
\begin{align}
\frac{\partial c_i}{\partial t} = D_i \Delta c_i + R(c_1, \ldots, c_n)
\end{align}
where $c_i$ denotes the concentration of species $i$, with diffusion constant $D_i$ and reaction term $R(c_1, \ldots, c_n)$. As network components we include the morphogens FGF10 (F10), FGF8 (F8), BMP, WNT and the structure AER with diffusion constant D  as well as the respective receptor-ligand-complexes with FGF10, denoted F10R, with FGF8, denoted F8R, with BMP, denoted BRa, and with WNT, denoted WFZ with diffusion constant DR. The reaction terms are set to
{
\footnotesize
\begin{align}
\begin{split}
R(F10) &=\rho_{F10}p_{FGF10} - d \, F10 \label{reac}\\   
&\quad+(1-\mathds{1}_{Mes})(k_{off} F10R-k_{onR} F10(RT_{F10}-F10R))\\ 
R(F8) &= \rho_{F8} p_{FGF8}-d \,F8\\  
& \quad -\mathds{1}_{Mes} k_{onR} F8 (RT_{F8}-F8R)+\mathds{1}_{Mes} k_{off} F8R\\  
R(WNT) &=\rho_{WNT} p_{WNT}-d\,  WNT   \\    
&\quad -k_{onR}  WNT (RT-WFZ)+k_{off} WFZ\\  
R(BMP) &= \rho_{BMP}p_{BMP}-d\, BMP\\  
&\quad-k_{onR} BMP (RT_{BRa}-BRa)+k_{off} BRa\\
R(F10R) &=k_{onR} F10 (RT_{F10}-F10R)-(k_{off}+d_{F10R})F10R\\  
R(F8R) &= k_{onR} F8(RT_{F8}-F8R)-(k_{off}+d_{F8R})F8R\\  
R(WFZ) &=k_{onR} WNT(RT-WFZ)-(k_{off}+d_{BR})WFZ   \\  
R(BRa) &= k_{onR} BMP(RT_{BRa}-BRa)-(k_{off}+d_{BR})BRa  \\
R(AER) &= \rho_{AER}\left( \mathds{1}_{AER}\textstyle\frac{WFZ^2}{WFZ^2+K_{WNT}^2}\textstyle\frac{BRa^2}{BRa^2+K_{BRa\_AER}^2}\right)\\  
& \quad-d_{AER}\textstyle\frac{K_{WNT1}^2}{WFZ^2+K_{WNT1}^2}\textstyle\frac{K_{BRa\_AER1}^2}{BRa^2+K_{BRa\_AER1}^2}AER
\end{split}
\end{align}}
with production terms{
\footnotesize
\begin{align}
\begin{split}
p_{FGF10} &= \textstyle\frac{1}{20} \mathds{1}_{lowMes}+\textstyle\frac{F8R^2}{F8R^2+K_{F8R\_F10}^2}\mathds{1}_{Mes}\\  
p_{FGF8 }&=\mathds{1}_{AER} AER  \textstyle\frac{WFZ^2}{WFZ^2+K_{WNT\_FGF}^2}\textstyle\frac{BRa^2}{BRa^2+K_{BRa\_FGF}^2}\textstyle\frac{K_{BRi}^2}{BRa^2+K_{BRi}^2} \\  
p_{WNT} &=\textstyle\frac{F10R^2}{F10R^2+K_{F10R}^2}(\mathds{1}_{Ect}+AER\mathds{1}_{AER}) \\  
p_{BMP} &= \left(\mathds{1}_{Mes}\textstyle\frac{F8R^2}{F8R^2+K_{F8R}^2}+(1-\mathds{1}_{Mes})\textstyle\frac{WFZ^2}{WFZ^2+K_{WNT}^2}\right)\textstyle\frac{K_{BRi\_Bmp}^2}{BRa^2+K_{BRi\_Bmp}^2}.    
\label{prod}
\end{split}
\end{align}}

The indicator function $\mathds{1}_{X}$ denotes that the corresponding reaction only take place in domain X, as the expression of the morphogens and receptors is restricted to different parts of the limb bud domain. Here, 'lowMes' denotes the lower half of the Mesenchyme, 'Mes' the Mesenchyme, 'Ect' the Ectoderm, and 'AER' the AER. To account for the inhibitory and activating effects on gene expression as displayed in the regulatory network (Fig.\ref{intro}b) the model uses Hill kinetics with maximum production rates $\rho_c$ and Hill constants $K_{xx}$.

\subsection{Parameter Inference}
\begin{figure}[!t]
\centering
\includegraphics[width=\columnwidth]{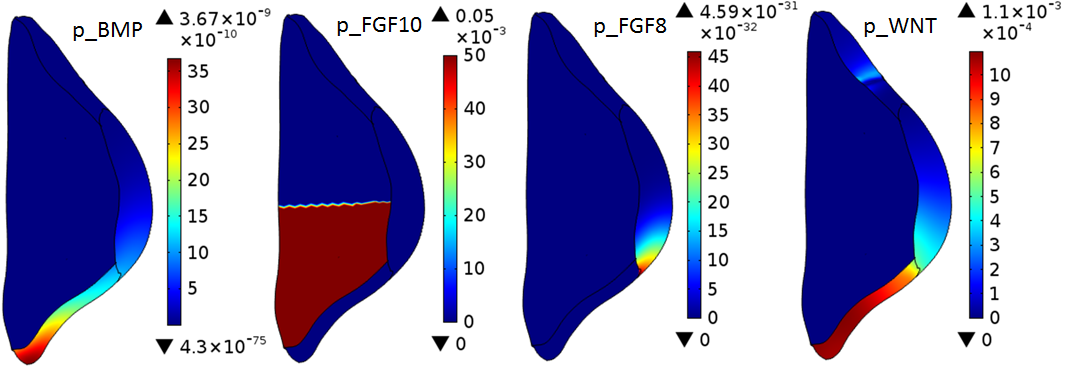} 
\caption{Predicted expression patterns with the initial parameter values. }
\label{start}
\end{figure}

To prepare for the inference of the parameter values, $\theta$, we first generate \textit{in silico} data with the parameter set, $\theta_0$, in Table 1. These result in the expression patterns in Fig. \ref{intro}c. As starting values for parameter inference we then use diffusion constants (in $[\mu m^2 h^{-1}]$) in Table 1, as their physiological range is typically rather well known. For the other parameters we set all maximal production rate to 2, all Hill constants and maximum receptor capacities to 1, and all binding and degradation rates to 0.01. These initial parameter values result in the expressions patterns shown in Fig. \ref{start}. Starting from those, we aim at finding a set of parameter values, with which we can reproduce the expression patterns in Fig. \ref{intro}c.

Quantitative, spatial expression data is currently not available in the limb bud. Therefore, the model only needs to match the observed patterns qualitatively. To quantify the goodness-of-fit of the simulated expression patterns, we used a mathematical formulation for the constraints that describe the desired patterns. Thus, based on the image data we require $p_{FGF10}$ to be present in the mesenchyme with a gradient, $p_{FGF8}$ to be uniformly present in AER, $p_{BMP}$ to be uniformly present in the ectoderm and AER as well as in the mesenchyme with a gradient and $p_{WNT}$ to be uniformly present in the ectoderm. The absence of WNT and F8 production in the mesenchyme and (in case of F8) in the ectoderm is already hard-coded by the indicator-functions in the PDEs. As the production rates range from 0 to 1 we choose by eye 0.3 as a threshold for presence, and a penalty term enforces presence of substance c in domain X, i.e.
\begin{equation}
\int_X{\max(0.3-c,0)^2 d\mu}.
\end{equation}
For uniform production, we added the penalty term
\begin{equation}
\int_X{(c-\max_X(c))^2 d\mu}.
\end{equation}
The constraint given by a gradient was approximated using a smoothed step function $\varphi_c(x)$ in x-direction, i.e.
\begin{equation}
\int_X{(0.3-c)_+^2  \varphi_c(x) +(c-0.1)_+^2(1-\varphi_c(x)) d\mu}.
\end{equation}
Adding all these constraints provides us with a fitness function $f(\theta)$ as a measure for the goodness-of-fit. For the \textit{in silico} generated expression pattern (Table 1) its value is \mbox{$f(\theta_{0})=3852$}, whereas for the initial parameter values in our optimisation we obtain \mbox{$f(\theta_{start})=13989$}. Only the relative, but not the absolute value of $f(\theta)$ matters.

\begin{table}[!t]
\caption{Parameter set $\theta_0$ to generate \textit{in silico} data} 
\centering 
\begin{tabular}{lcr} 
\hline\hline 
Name& Value& Description \\ 
\hline 
$D$ & 3600 & Diffusion constant of proteins \\ 
$DR$ & 36&Diffusion constant of receptors \\
$\rho_{F8}$ & 0.9& max. production rate of FGF8 \\
$\rho_{F10}$ & 7.2& max. production rate of FGF10 \\
$\rho_{BMP}$ & 1.8& max. production rate of BMP \\
$\rho_{WNT}$ & 1.8& max. production rate of WNT \\
$\rho_{AER}$ & 0.36& max. production rate of AER \\
$init_{F10}$ & 0& initial concentration of FGF10 \\
$d$ & 0.0036& degradation rate of morphogenes \\
$d_{AER}$ & 3.6& degradation rate of AER \\
$d_{BR}$ & 1.8& degradation rate of WFZ, BRa \\
$d_{F10R}$ & 0.36& degradation rate of F10R \\
$d_{F8R}$ & 0.288& degradation rate of F8R \\
$k_{onR}$ & 3.6 & on-binding rate to receptors\\
$k_{off}$ & 0.0036 & off-binding rate from receptors\\
$RT$ & 10 & max. capacity of WNT receptor\\
$RT_{F8}$ & 2.5 & max. capacity of FGF8 receptor\\
$RT_{F10}$ & 2.5 & max. capacity of FGF10 receptor\\
$RT_{BRa}$ & 0.6 & max. capacity of BMP receptor\\
$K_{F8R}$& 0.1 & Hill constant\\
$K_{F8R\_F10}$& 0.025 & Hill constant\\
$K_{F10R}$& 1 & Hill constant\\
$K_{BRa}$& 0.3 & Hill constant\\
$K_{BRa\_FGF}$& 0.01 & Hill constant\\
$K_{BRa\_AER}$& 0.01 & Hill constant\\
$K_{BRa\_AER1}$& 0.001 & Hill constant\\
$K_{BRi}$& 0.25 & Hill constant\\
$K_{BRi\_Bmp}$& 0.5 & Hill constant\\
$K_{WNT}$& 0.1 & Hill constant\\
$K_{WNT\_FGF}$& 0.1 & Hill constant\\
$K_{WNT1}$& 0.001 & Hill constant\\
$K_{WNT\_Bmp}$& 0.01 & Hill constant\\
\hline 
\end{tabular}
\label{tab:hresult}
\end{table}

A sensitivity analysis revealed that the expression patterns are sensitive to all parameters except for $K\_BRa\_AER$ and $K\_BRa\_AER1$. The parameters in the model are highly correlated and the vast dimension of the parameter space as well as the non-smooth fitness function render standard optimization methods such as Coordinate Search, gradient-based methods like SNOPT adjoint method, or random algorithms such as the Particle Swarm Optimization unable to improve the value of the fitness function and to arrive at a parameter set for which the model reproduces the patterns in Fig. \ref{intro}c. We note that the repeated simulation of the PDEs is computationally very expensive.

\subsection{A decision tree approach for the sequential inference of parameter values}
To overcome the difficulties described above, we developed a decision tree that also uses prior knowledge and that optimises parameter values sequentially given that most parameter values affect the patterning process only at certain time intervals. To determine the sequence, in which parameters are identified, we start with the initial conditions and the constitutive production rates and check which production rates are controlled by those factors that are initially present, here FGF10 production in the lower part of the mesenchyme. All production rates except for the direct downstream targets are set to zero. For each of these steps only a subset of the original parameters has a direct influence. This subset is tuned with help of a decision tree that checks whether the currently considered components are present in the correct domains. If all important features are reproduced, the algorithm goes on to check the next component activated downstream; if not, the parameters which have direct influence on the considered feature are doubled or halved depending on the sign of their influence and then the procedure of the decision tree in the current step is repeated until the constraint is fulfilled or a maximum number of calls is reached. For the model considered here this approach results in the steps
\begin{compactenum}
\item F10 production $\rightarrow$ check F10, F10R
\item F10 and WNT production $\rightarrow$ check WNT, WFZ in Ectoderm (Ecto)
\item F10, WNT, BMP, AER production $\rightarrow$ check BMP, BRa in Ecto, check AER, check WNT, WFZ in Ecto
\item F10, WNT, BMP, AER, F8 production$\rightarrow$ check F8, F8R, check BMP in Mesenchyme, check F10 in Mesenchyme
\end{compactenum}
The decision tree for the second step of the algorithm is illustrated in Fig. \ref{dec}. Starting with the parameter values $\theta_{start}$ results in a set of parameters $\theta_{dec}$, for which the model yields expression patterns (Fig.  \ref{dec_res}) close to the original patterns (Fig. 1c), and the fitness function $f$ reduces from \mbox{$f(\theta_{start})=13989$} to \mbox{$f(\theta_{dec})=5598$}.

 \tikzstyle{level 1}=[level distance=0.5cm, sibling distance=2cm]
\tikzstyle{level 2}=[level distance=0.6cm, sibling distance=2cm]
\tikzstyle{level 3}=[level distance=0.5cm, sibling distance=2cm]
\tikzstyle{level 4}=[level distance=0.6cm, sibling distance=2cm]
\tikzstyle{level 5}=[level distance=0.5cm, sibling distance=2cm]
\tikzstyle{level 6}=[level distance=0.7cm, sibling distance=2cm]
\tikzstyle{level 7}=[level distance=0.5cm, sibling distance=2cm]
\tikzstyle{level 8}=[level distance=0.4cm, sibling distance=2cm]

\tikzstyle{quest} = [ellipse, text width=6em, text centered, draw]
\tikzstyle{instr} = [rectangle, text width=8em, text centered, draw]
\tikzstyle{ans} = [minimum width=8pt, text centered, inner sep=0pt]
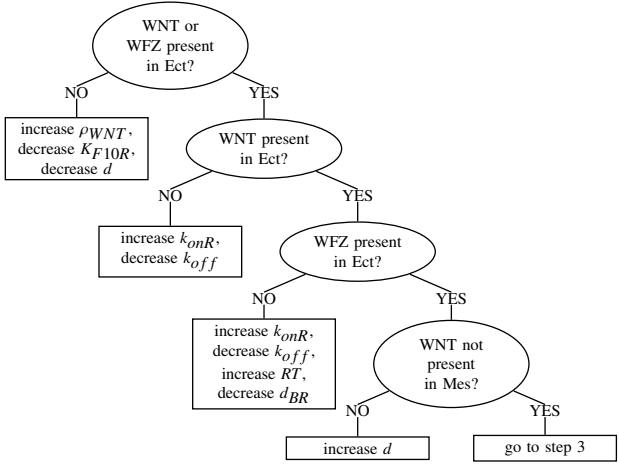
\begin{figure}[!ht]
\resizebox{\linewidth}{!}{
 \begin{tikzpicture}[grow=south,child anchor=north]
\tiny
\node[quest]{WNT or WFZ present in Ect?}
child {
    node[ans]{NO}
    child {
        node[instr]{increase $\rho_{WNT}$, decrease $K_{F10R}$, decrease $d$}
            }}
    child {
    node[ans]{YES}
    child {
            node[quest]{WNT present in Ect?}
            child {
    node[ans]{NO}
    child {
        node[instr]{increase $k_{onR}$, decrease $k_{off}$}
		            }}
child {
    node[ans]{YES}
    child {
        node[quest]{WFZ present in Ect?}
            child{node[ans]{NO}
            child{ node[instr]{increase $k_{onR}$, decrease $k_{off}$, increase $RT$, decrease $d_{BR}$}
}
                     }       
                 child {
    node[ans]{YES}
    child {
        node[quest]{WNT not present in Mes?}
        child{node[ans]{NO}
    child {
        node[instr]{increase $d$}
		            }}
		            child{node[ans]{YES}
    child {
        node[instr]{go to step 3}
		            }}
                    }
            }
            }
            }
            }
            };
 \end{tikzpicture}}

 \caption{The decision tree performed by the parameter inference algorithm in step 2.}
 \ \label{dec}
 \end{figure}

\subsection{Local optimization with SNOPT}
The parameter values recovered from the decision tree approach are used as initial values to perform local optimization using the SNOPT algorithm implemented in COMSOL 4.4, a gradient-based algorithm that uses sequential quadratic programming (SQP) methods and which  requires the least computation time for this problem among the COMSOL methods.  To calculate the gradient, the adjoint method was used, as it is most efficient in case of many parameters.  At this step of our approach we compare the optimised patterns (Fig. \ref{opt_expr}) to the \textit{in silico} generated patterns (Fig. \ref{intro}c) directly, i.e. as objective function $J(\theta)$ we integrate the difference between the \textit{in silico} generated expression values and the model output, while adding scaling factors $\lambda_i$ as additional parameters as we aim at reproducing the patterns and no absolute values, i.e. $J(\theta)=$
{\footnotesize
\begin{align}
\begin{split}
\int_{AER}{(p_{FGF8}-\lambda_1 \hat{p}_{FGF8})^2 \,d\mu} &+  \int_{AER, Ect}{(p_{WNT}-\lambda_2 \hat{p}_{WNT})^2\,d\mu} \\
+ \int_{all}{(p_{BMP}-\lambda_3 \hat{p}_{BMP})^2 \,d\mu}&+  \int_{all}{(p_{FGF10}-\lambda_4 \hat{p}_{FGF10})^2 \,d\mu }
\end{split} \nonumber
\end{align}}
This formulation allows us to evaluate the recovered parameter values with statistical measures, i.e. calculate the profile likelihood. To reduce computational costs we exclude the Hill constants, which are correlated in particular with expression and decay rates, from the optimization, and we reduce the accuracy of the simulation. Thereby, we can reduce the computation time from 24 hours to approx. 2 hours for one optimization. Within 100 iterations the algorithm halves the value of the objective function and recovers parameter values $\theta_{opt}$ with a fitness function value $f(\theta_{opt})=4526$ (Fig. \ref{opt_expr}), reasonably close to the fitness function value of the target pattern, \mbox{$f(\theta_{0})=3852$} (Fig. \ref{intro}c).

 \begin{figure}[!t]
\centering
\includegraphics[width=\columnwidth]{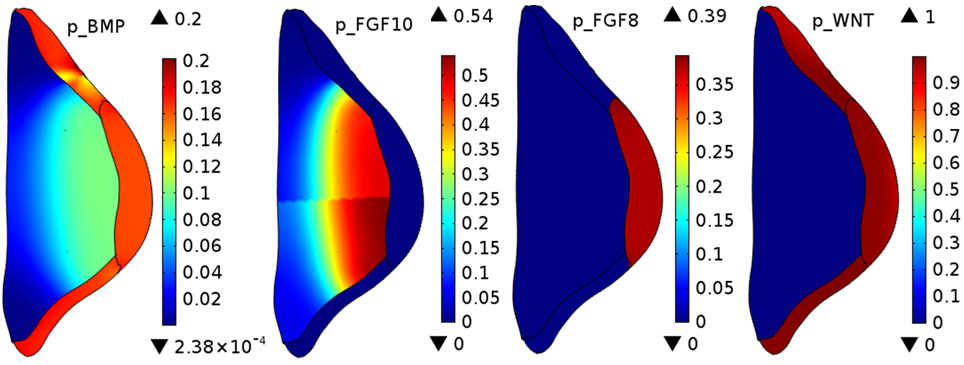} 
\caption{Expression patterns after the decision-tree based parameter inference.}
\label{dec_res}
\end{figure}

 \begin{figure}[!t]
\centering
\includegraphics[width=\columnwidth]{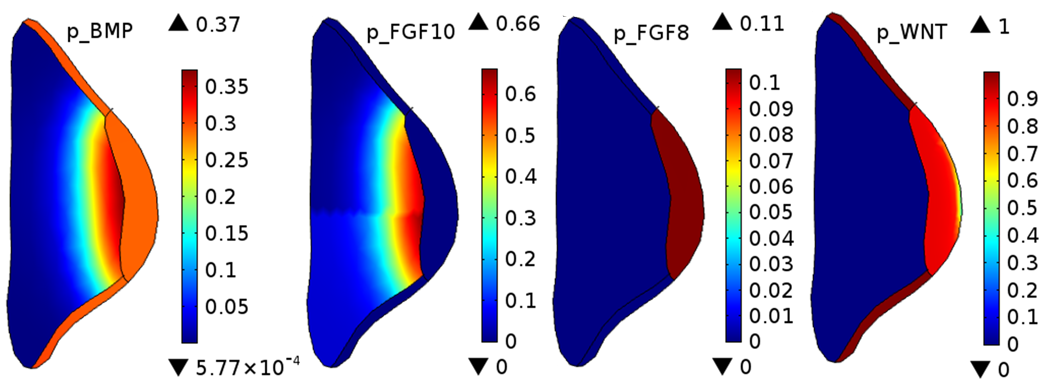} 
\caption{Expression patterns after the  optimization with SNOPT, using the parameters recovered by the decision tree algorithm as initial values.}
\label{opt_expr}
\end{figure} 

\section{Methods}
The model was solved with COMSOL Multiphysics using the Optimization Toolbox (version 4.4) and the MATLAB Livelink (version 4.3b) as described before \cite{Germann:ProceedingsOfComsolConference2011:2011, Menshykau:ProceedingsOfComsolConference2013Rotterdam:2013}.

\section{Acknowledgements}
We thank Z. Karimaddini, and R. Zeller for discussions. The work was in part funded by the SNF Sinergia and NeuroStemX grants.

\bibliographystyle{ieeetr}

\begin{thebibliography}{1}

\bibitem{Iber:CurrOpinGenetDev:2012}
D.~Iber and R.~Zeller, ``Making sense-data-based simulations of vertebrate limb
  development.,'' {\em Curr Opin Genet Dev}, vol.~22, pp.~570--577, 12 2012.

\bibitem{Probst:Development:2011}
S.~Probst, C.~Kraemer, P.~Demougin, R.~Sheth, G.~R. Martin, H.~Shiratori,
  H.~Hamada, D.~Iber, R.~Zeller, and A.~Zuniga, ``Shh propagates distal limb
  bud development by enhancing cyp26b1-mediated retinoic acid clearance via
  aer-fgf signalling.,'' {\em Development}, vol.~138, pp.~1913--1923, 5 2011.

\bibitem{Zeller:NatRevGenet:2009}
R.~Zeller, J.~Lopez-Rios, and A.~Zuniga, ``Vertebrate limb bud development:
  moving towards integrative analysis of organogenesis.,'' {\em Nat Rev Genet},
  vol.~10, pp.~845--858, 12 2009.

\bibitem{Germann:ProceedingsOfComsolConference2011:2011}
P.~Germann, D.~Menshykau, S.~Tanaka, and D.~Iber, ``Simulating organogensis in
  comsol,'' in {\em Proceedings of COMSOL Conference 2011}, 9 2011.

\bibitem{Menshykau:ProceedingsOfComsolConference2013Rotterdam:2013}
D.~Menshykau, A.~Shrivastsan, P.~Germann, L.~Lemereux, and D.~Iber,
  ``Simulating organogenesis in comsol: Parameter optimization for pde-based
  models.,'' in {\em Proceedings of COMSOL Conference 2013, Rotterdam}, 10
  2013.

\end{thebibliography}

\end{document}